# Study on the Best Uses of Technology in Support of Project-Based Learning

James Taylor
Georgia Institute of Technology
Atlanta, Georgia, USA
Jtaylor323@gatech.edu

**ABSTRACT**

Project-Based Learning (PBL) is a teaching technique in which authentic, real-world projects are used as the primary vehicle to drive the student's learning experience. This technique has been found to be very effective, but its overall adoption rate is relatively low, in part due to teachers' unfamiliarity with how to best use technology to successfully implement it. This research study involved a comprehensive survey of supportive technology tools, as well as secondary survey research from students and teachers with actual experience in PBL. The goal was to determine which types of technology tools were most supportive of PBL. Overall, the study found that teachers and students are mostly aligned with regards to the importance of technology and the effectiveness of various types of tools. Tools which fostered collaboration amongst teacher and students were ultimately deemed the most effective, but content-development and assessment tools were also found to be particularly helpful.

**OVERVIEW**

Project-based learning (PBL) is a pedagogical technique in which students, often working in teams, may combine a variety of subject matter disciplines or skills as they work to solve authentic, real-world problems. [1] In contrast to "final projects" given to students after a period of more traditional lecture and textbook-based instruction, with PBL the projects are started at the beginning of the course, and in fact represent the core focus of the curriculum, in that the project is used as the primary vehicle to drive the student's learning experience. [12]

PBL allows students to engage with course subjects in a manner that encourages teamwork, with an emphasis on communication, collaboration and critical thinking skills. [14] Additionally, PBL effectively "flips the classroom" by allowing teachers to take on more of a facilitative role in the student's learning, providing support when students get stuck, and providing oversight to ensure students stay on track. [1]

Technology is a key supportive component to this teaching method. Students can use technology to research topics, and then to create digital artifacts like websites, blogs and podcasts with the results of their findings. This allows students to embrace a topic using the same kinds of technologies they are used to consuming outside of the classroom. Collaborative technology also enables students to continue their work with fellow students (and teachers) after they leave the physical classroom. Finally, technology gives teachers the ability to monitor students' progress more readily, and to provide individualized feedback. [3]

Numerous studies have found PBL to offer clear benefits to students, particularly in the ability to develop students' critical thinking skills and to reinforce how the concepts they are learning can be applied to the real world. [12] PBL has also been cited as a particularly effective teaching technique for at-risk students. [4] Still, only a small number of schools (around one percent) have formally adopted this approach to learning. One major reason for the low rate of adoption is that teachers often lack the support they need to make the transition from traditional lecture-based learning to PBL. This is particularly true with regards to access to and understanding of the appropriate supportive technology that can help both the teacher and student succeed in this new teaching approach. [11]

Understanding that teachers embracing PBL may need more guidance on which types of supportive technology are most helpful in the successful implementation of this teaching method, this study aimed to answer the following questions:

- Which types of technological tools do teachers find to be most helpful in support of their Project-Based Learning efforts?
- Which types of technological tools used in support of PBL do students find to be most helpful and engaging?
- Is there alignment between the types of tools that teachers find effective and students find interesting?
- For which PBL-specific teaching practices does technology provide the greatest support?
- Overall, how integral is technology to the delivery of PBL?

**RESEARCH APPROACH AND METHODOLOGY**

The research approach for this study consisted of both primary and secondary research. The primary research focused on a broad analysis and categorization of currently available supportive technology tools. Secondary research consisted of two surveys. The first survey was conducted with teachers who had experience in the delivery of PBL; the goal of this survey was to determine which types of technology tools these teachers found most helpful in their implementation of PBL. The second survey was conducted with students who had taken at least one class in which PBL was used as a primary teaching technique; the goal of this second survey was to find out which types of technology tools these students found most effective and engaging.

**SURVEY OF AVAILABLE TECHNOLOGY TOOLS IN SUPPORT OF PBL**

Background research [2, 6, 8, 9, 10, 13] uncovered a large and diverse set of available technological tools in support of PBL – so many so that simply listing each tool on a survey would have been unwieldy. Consequently, it was necessary to perform a detailed analysis of the available tools. The goal of this analysis was to uncover some higher-level categories that could then be referenced in the forthcoming surveys.

Each tool was reviewed and logged in a catalog. Details logged included:

- The name of the product and its vendor
- A brief description of the product
- A brief description of how the product could be used to support PBL
- If the product was perceived to be a market leader
- Whether use of the product was free or incurred a cost

This catalog, which ultimately contained nearly 70 technology tools, yielded the following set of high-level categories:

- **Assessment Tools**: These tools allow teachers to determine the extent to which students are grasping core concepts, as well as gathering feedback from students. Popular examples: Edmodo Snapshot, Google Forms/Surveys
- **Brainstorming Tools**: These tools enable students to throw out a variety of ideas during brainstorming sessions, then look for ways in which ideas are connected. Includes concept mapping and note taking tools. Popular examples: SimpleMind+, Mindmeister, Evernote
- **Collaboration Tools**: These tools foster collaboration between teachers and students, allowing for "over the shoulder" reviews of content and enabling peer feedback. Popular examples: Google Docs, Microsoft Office Online, Edmodo, Skype, Slack
- **Content Development Tools for Students**: These tools enable students to create content (e.g. presentations, movies, spreadsheets, papers, collages, blogs, podcasts) in support of their projects. Popular examples: Google Docs, Glogster, Prezi, Storify, Dipity. AudioBoom, PowerPoint
- **Knowledge Transfer Tools for Teachers**: These tools enable teachers to create presentations and whiteboards, as well as provide curated content which students can use in support of projects. Popular examples: Newsela, Screencast-o-Matic, Camtasia
- **Online Libraries of Project Ideas**: These tools provide teachers with the ability to search libraries of existing project ideas, often focusing the search on a particular subject matter or age range. Popular examples: CraftED Curriculum, GlobalSchoolNet.org, Educurious, NextLesson
- **PBL Platforms/Frameworks**: These are platforms which teachers can use to manage all aspects of their PBL delivery, including built-in collaboration, planning and assessment tools. Popular examples: Project Foundry, Novare PBL Platform, CrowdSchool, Project Foundry
- **Planning Tools**: These include online planning forms specific to PBL, time management tools and teacher/student messaging tools. Popular examples: Buck Institute for Education's Online Planning Forms, FlexTime Manager, Remind
- **Standards Compliance Tools**: These tools help teachers locate sample projects and content that support specific Common Core standards. Popular examples: NextLesson, Educurious, Newsela

Because Content Development is such a core component of PBL, additional analysis was conducted to identify a core set of content types that could be created with the identified Content Development Tools for Students:

- Blogs and Stories
- Collages and Scrapbooks
- Documents
- Drawings and Diagrams
- Interactive Slideshows
- Movies
- Podcasts
- Presentations
- Social Media
- Spreadsheets / Data Analysis
- Timelines

Finally, each tool was mapped to one or more of the seven core teaching practices that the Buck Institute for Education (BIE) has noted as key to a successful PBL implementation (Figure 1). The goal of this mapping was to gather some additional insight into which teaching practices these technology tools may be most supportive of.

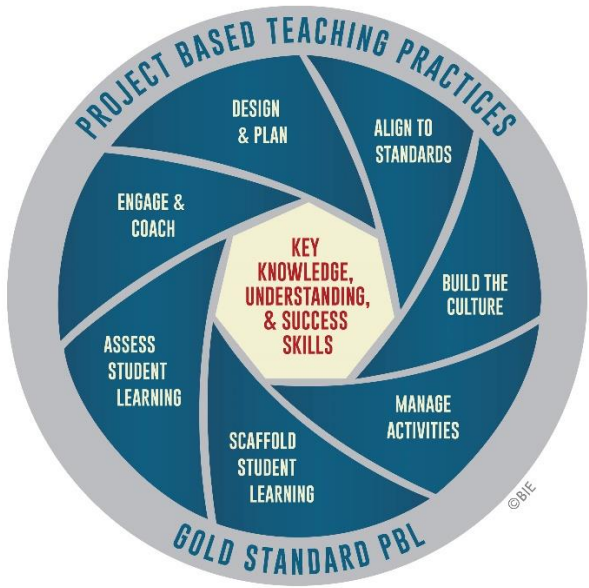

**Figure 1: The Seven Core Project-Based Learning Teaching Practices [7]**

## TEACHER SURVEY

### Overview

This survey was designed to reach teachers with experience in the delivery of Project-Based Learning. A series of questions were designed to determine the types of technological tools and content types these experienced teachers found to be most helpful. Additionally, the survey included some open-ended questions designed to gather more insight into specific tools teachers found particularly helpful (or unhelpful). Another set of questions gauged teachers' attitudes to the importance of technology relative to the success of PBL, as well as to determine the extent to which cost was a factor in the adoption of certain tools. Finally, a few general questions were included to classify the respondents based on years of experience and grade level of students taught.

### Survey Methodology

This web-based survey was administered via Georgia Tech's Qualtrics platform. Participants were recruited through word of mouth, as well as in response to some Facebook and Twitter-based advertising campaigns targeting individuals with interest in Project-Based Learning. As an incentive, prizes of $25 Amazon.com gift cards were offered to three randomly-chosen survey participants.

The survey ultimately yielded 23 responses. Respondents had an average of 6.76 years' experience in the delivery of Project-Based Learning, with experience somewhat evenly distributed across a variety of grade levels (Figure 2).

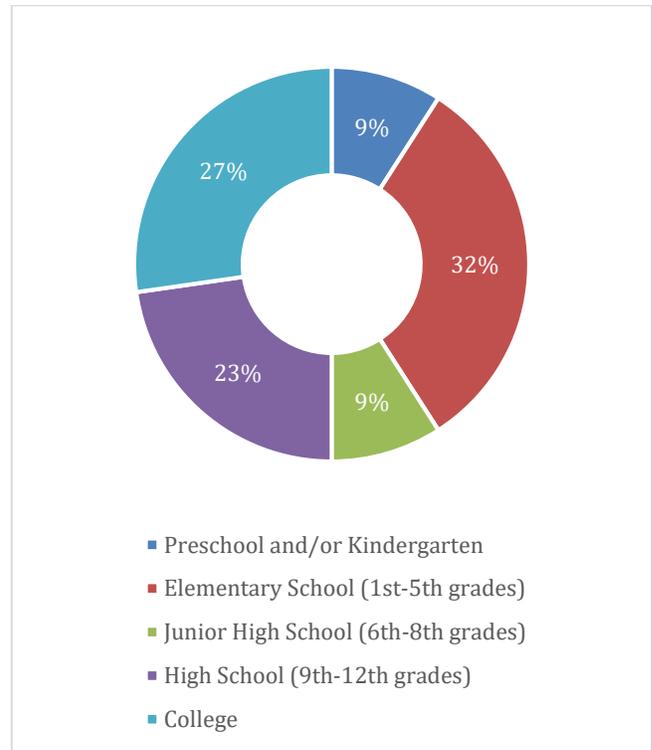

**Figure 2: Grade levels in which teacher respondents had used PBL**

### Key Insights from Teacher Survey

Background research [11] had indicated that teachers new to PBL often struggled to find information on the various types of technology tools in support of PBL. Interestingly, this was not a significant issue amongst survey respondents; overall, nearly 83% or respondents felt they had access to adequate information, while only 4% of respondents felt their access to information was inadequate (Figure 3).

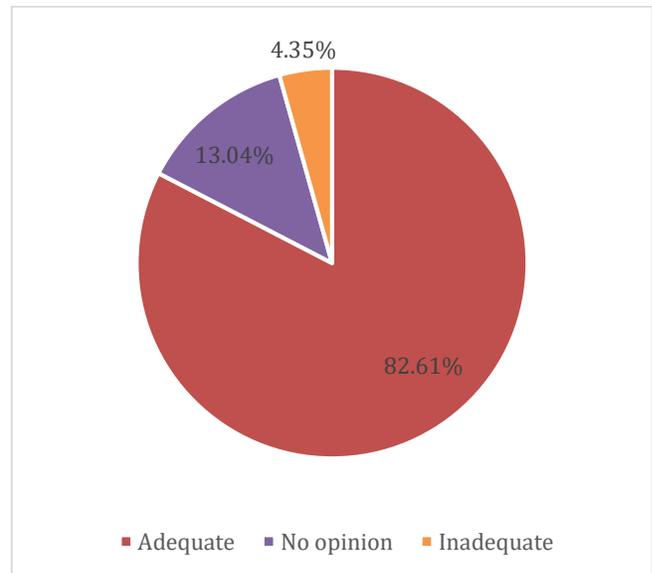

**Figure 3: Teacher respondents' reported level of access to information on PBL-supportive technology tools**

Further segmenting the data by years of experience with PBL, there was only a 5% difference in "adequate" responses between teachers with less than five years of experience and teachers with five or more years of experience. This may indicate that access to this information has improved in more recent years, or that it was not as big an issue as the referenced background research made it out to be.

That said, a follow-up question, asked teachers which professional development tools provided the greatest benefit in improving their delivery of PBL. These results (Figure 4) indicate that teachers still have an interest in learning how to better use technology, as well as how to build the technical skills (e.g. video production) that can help them facilitate more effectively.

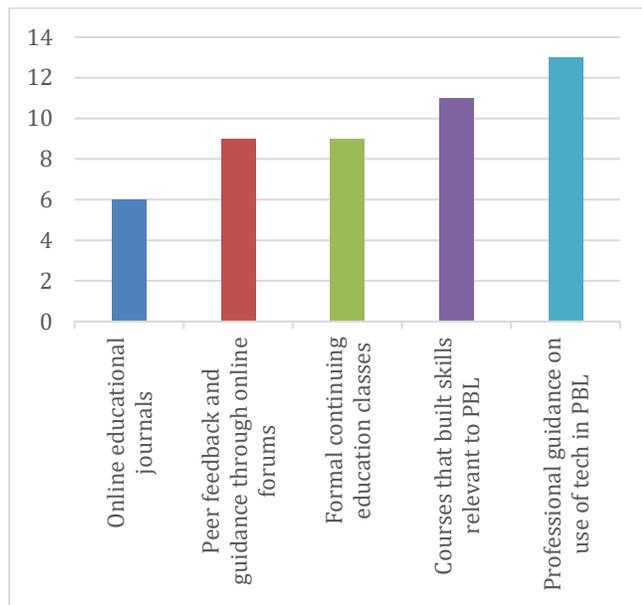

**Figure 4: Professional development techniques teacher respondents found most helpful in building PBL skills**

Getting into the details of the technology tools, the respondents were provided with the list of nine technology tool categories, then were asked two questions. The first question queried as to the frequency of usage for each type of tool (using a five-point scale), while the second question asked teachers to rate the tools from most effective to least effective. These results were then averaged and sorted by highest to lowest score (Table 1).

| Technology Type | Greatest Impact | Most Used | Average |
|---|---|---|---|
| Collaboration Tools | 0.85 | 0.82 | 0.84 |
| Assessment Tools | 0.76 | 0.64 | 0.70 |
| Content Development Tools for Students | 0.65 | 0.73 | 0.69 |
| Knowledge Transfer Tools for Teachers | 0.61 | 0.60 | 0.61 |
| Brainstorming Tools | 0.59 | 0.43 | 0.51 |
| Planning Tools | 0.49 | 0.40 | 0.44 |
| Online Libraries of Project Ideas | 0.47 | 0.40 | 0.43 |
| PBL Platforms/Frameworks | 0.38 | 0.36 | 0.37 |
| Standards Compliance Tools | 0.21 | 0.40 | 0.30 |

**Table 1: Teacher rankings of technology tool types**

A follow-up question asked teachers to indicate (using a five-point scale) how often students used various types of content in fulfillment of their projects, and whether teachers found those types of content to be particularly effective.

| Content Type | Used | Effective | Average |
|---|---|---|---|
| Documents | 0.88 | 0.94 | 0.91 |
| Drawings and Diagrams | 0.78 | 0.89 | 0.83 |
| Interactive Slideshows | 0.72 | 0.89 | 0.81 |
| Movies | 0.68 | 0.83 | 0.76 |
| Presentations | 0.66 | 0.83 | 0.74 |
| Spreadsheets / Data Analysis | 0.59 | 0.83 | 0.71 |
| Blogs and Stories | 0.56 | 0.78 | 0.67 |
| Timelines | 0.49 | 0.78 | 0.63 |
| Podcasts | 0.49 | 0.72 | 0.61 |
| Social Media | 0.40 | 0.61 | 0.51 |
| Collages and Scrapbooks | 0.39 | 0.44 | 0.42 |

**Table 2: Teacher rankings of project content types**

These results (Table 2) imply that students tend to gravitate towards more traditional documents and diagrams when preparing content in fulfillment of their projects, although more dynamic types of content like interactive slideshows and movies were also found to be particularly effective.

## STUDENT SURVEY

### Overview

This survey was designed to reach students who had taken at least one course in which Project-Based Learning had been used as the primary pedagogical method. The main purpose of the survey was to gather information about what types of technological tools students had used when working on their projects, as well as which types of tools they found most (and least) effective. The survey also included some general questions regarding the level of usage and interest with several types of content-creation tools. Some final open-ended questions were designed to gather additional insights into specific tools students found helpful (or unhelpful), as well as to learn more about students' overall attitudes towards the use of technology in completion of their projects.

**Survey Methodology**

This web-based survey was administered via Georgia Tech's Qualtrics platform and was offered to students in the Fall 2017 session of Georgia Tech's Educational Technology course; these students were targeted in part because this course is built around PBL, thus providing a group of students with at least nominal experience with this type of teaching technique. Survey participants were recruited via the class' online Piazza forum, and three participation tokens were offered to survey participants.

This survey ultimately gathered 52 responses. Over 75% of survey participants had taken at least one additional course in which PBL was the core teaching method, indicating that this group of participants had solid experience in this teaching method (Figure 5).

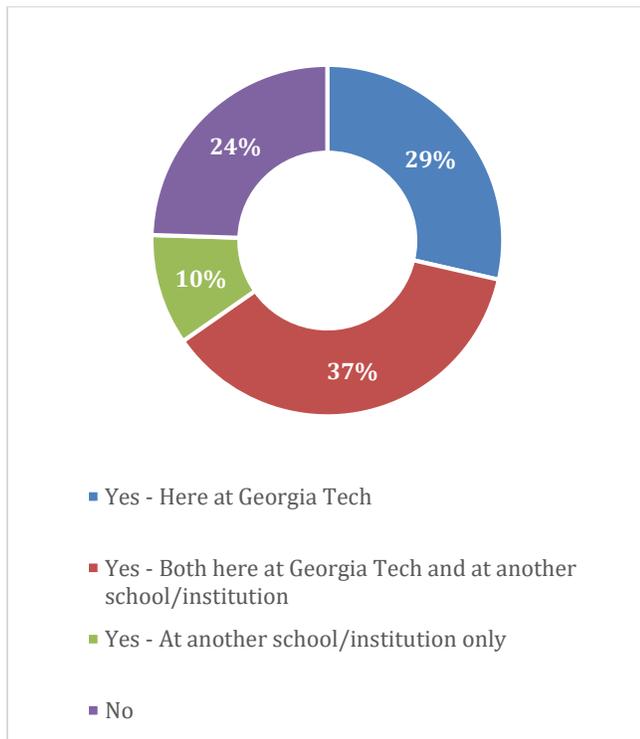

**Figure 5: Prior PBL experience amongst student respondents**

**Key Insights from Student Survey**

Student respondents were provided with the list of seven[1] technology tool categories, then were asked two questions. The first question queried as to the frequency of usage for each type of tool (using a five-point scale), while the second question asked students to rate the tools from most effective to least effective. These results were then averaged and sorted from highest to lowest score (Table 3).

| Technology Type | Most Helpful | Most Used | Student Average | Teacher Average |
|---|---|---|---|---|
| Collaboration Tools | 0.73 | 0.78 | 0.75 | 0.84 |
| Content Development Tools for Students | 0.62 | 0.68 | 0.65 | 0.69 |
| Assessment Tools | 0.61 | 0.66 | 0.64 | 0.70 |
| Knowledge Transfer Tools for Teachers | 0.50 | 0.71 | 0.60 | 0.61 |
| Planning Tools | 0.54 | 0.60 | 0.57 | 0.44 |
| Online Libraries of Project Ideas | 0.45 | 0.60 | 0.53 | 0.43 |
| Brainstorming Tools | 0.55 | 0.43 | 0.49 | 0.51 |

**Table 3: Student rankings of technology tool types (with teacher rankings also supplied for comparison)**

A follow-up question asked students to rank (on a five-point scale) how often they created various types of content in support of their projects, and which of those content types they found interesting.

| Content Type | Used | Interest | Student Average | Teacher Average |
|---|---|---|---|---|
| Drawings and Diagrams | 0.79 | 0.93 | 0.86 | 0.83 |
| Documents | 0.90 | 0.70 | 0.80 | 0.91 |
| Spreadsheets / Data Analysis | 0.74 | 0.81 | 0.78 | 0.71 |
| Presentations | 0.76 | 0.65 | 0.71 | 0.74 |
| Interactive Slide shows | 0.64 | 0.70 | 0.67 | 0.81 |
| Time lines | 0.64 | 0.56 | 0.60 | 0.63 |
| Blogs and Stories | 0.47 | 0.70 | 0.59 | 0.67 |
| Social Media | 0.54 | 0.60 | 0.57 | 0.51 |
| Movies | 0.52 | 0.63 | 0.57 | 0.76 |
| Podcasts | 0.42 | 0.49 | 0.45 | 0.61 |
| Collages and Scrapbooks | 0.44 | 0.30 | 0.37 | 0.42 |

**Table 4: Student rankings of content types (with teacher rankings supplied for comparison)**

Overall, these results (Table 4) indicate that student and teacher respondents were closely aligned as to the relative importance of drawings, diagrams, documents and data in the documentation of their projects. Additionally, both students and teachers find interactive presentations and slide shows to be effective presentation tools.

---

[1] Two of the nine technology tool categories, PBL Platforms/Frameworks and Standards Compliance Tools, were deemed to be so teacher-specific that they didn't warrant inclusion in the student-targeted survey.

It's worth noting that some of the content types, such as podcasts, collages and scrapbooks, might be poorly suited for the types of projects that graduate-level Computer Science students are typically asked to complete. This might explain their relative lack of popularity amongst students versus teachers in this survey, and broadening the student survey to a more diverse population of students might yield some different results.

**KEY TAKEAWAYS**

Completion of the primary and secondary research for this study makes it possible to revisit the key research questions with new insights.

**Question 1: What types of technology tools do teachers find to be most helpful in support of Project-Based Learning?**

Collaboration Tools were far and away the number one choice by teachers, which is understandable given PBL's emphasis on group projects and on shifting the role of the teacher from lecturer to facilitator. Next on the list were Assessment Tools, which was telling since teachers often state that figuring out how to assess students' learning (and grade the corresponding projects) is one of the most challenging aspects of PBL. [5] A close third on the list were Content Development Tools for Students, which is not surprising given PBL's focus on students creating content as a vehicle for showing the results of their projects.

Last on the list were Standards Compliance Tools. It's worth noting that there was a nearly 20-point gap between the level of use and the perceived effectiveness for this tool type (see Table 1). This could indicate that teachers feel that standards have a limited impact on the success of their PBL efforts, even though they may be required to show at least nominal adherence to standards. It was also interesting that PBL Platforms and Frameworks ranked so low, particularly since these tools often position them to be "one stop shops" for PBL initiatives. This could indicate the respondents lack access to these more comprehensive delivery tools, or that they prefer the flexibility that a more loosely structured collection of tools can provide.

**Question 2: Which types of technology tools in support of PBL do students find to be most helpful and engaging?**

As with teachers, students found Collaboration Tools to be the most significant supportive technology. This was backed by a follow-up question which asked students for the names of specific tools they felt were helpful in completion of their projects. Mapping the responses back to the overall technology type found that 61% of students mentioned specific collaboration tools they found essential (Figure 6).

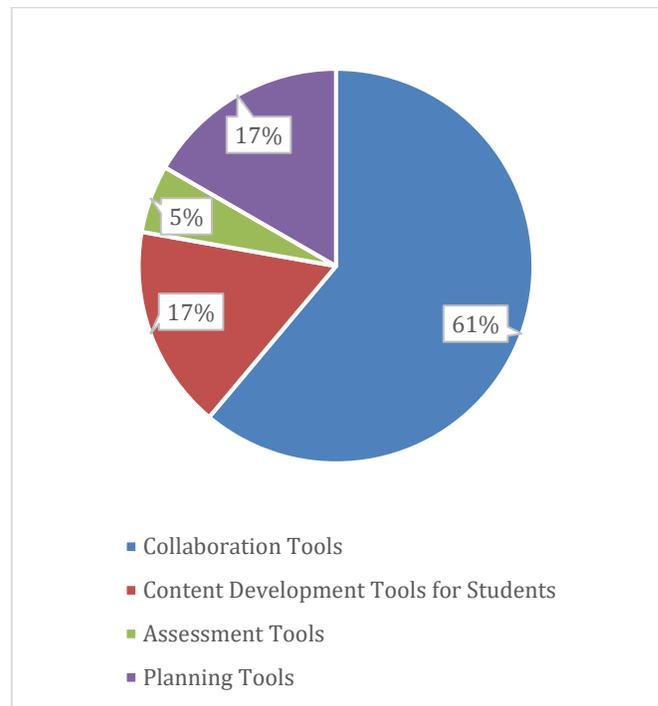

**Figure 6: Tool categories for anecdotally-cited student tools**

Students also found Assessment Tools, Planning Tools and Content Development Tools to be particularly helpful in creation of their projects.

**Question 3: Is there alignment between the types of tools that teachers find effective and students find interesting?**

In general, there was a very high degree of alignment in overall rankings amongst both teachers and students across the overall categories. The correlation coefficient between student interest and teacher effectiveness was 0.7884, indicating a relatively positive correlation between the two factors. Most notably, Collaboration Tools were the top-ranked technology tool for both students and teachers.

The major deviation between student and teacher rankings was with regards to Planning Tools, which students ranked higher. As one student noted in anecdotal comments, "if you are failing planning, you will most likely fail your project".

**Question 4: For which PBL-specific teaching practices does technology provide the greatest support?**

During primary research, each surveyed technology tool was mapped to one or more of the seven core PBL Teaching Practices identified by the Buck Institute for Education. Then, for each tool type category, a percentage was calculated based on the number of times each PBL Teaching Practice was mapped across all tools of that type. This percentage was then multiplied by a ranking factor (based on teachers' rankings of the overall technology tool types). The resulting scores were totaled across each PBL Teaching Practice (Table 5).

| Core PBL Teaching Practice | Score |
|---|---|
| Design & Plan | 2.36 |
| Engage & Coach | 2.36 |
| Build the Culture | 2.26 |
| Manage Activities | 1.80 |
| Assess Student Learning | 1.59 |
| Scaffold Student Learning | 0.74 |
| Align to Standards | 0.60 |

**Table 5: Technology relevance ranking of PBL-aligned teaching practices**

These results imply that while technology can help support all PBL Teaching Practices, it is most impactful in the following areas:

- Helping teachers to design and plan for the successful execution of their projects, particularly using planning guides and time management tools that allow them to make the most efficient use possible of precious classroom time
- Providing teachers with the ability remain in close contact with students throughout the project, as well as allowing students the ability to collaborate with each other, even when not in the same physical space
- Giving students the creative freedom to explore the topics underlying their projects in ways that are meaningful to them, while simultaneously ensuring they are equipped with the core knowledge needed to stay focused.

Collaboration, Assessment and Content Development tools are key to helping achieve these goals, and teachers who are new to PBL would be well advised to spend time investing in the skills needed to use these tools to their greatest effectiveness.

**Question 5: Overall, how integral is technology to the delivery of PBL?**

Nearly 74% of teachers and 63% of students thought that technology tools were either very important or extremely important relative to their successful delivery of Project-Based Learning. Thus, it was clear that amongst survey respondents at least, technology is considered an important tool, although one anecdotal comment from a teacher was particularly illuminating:

*"The most important thing to remember is that technology is **a** tool not **the** tool. Always have a plan B in the event that technology does not work. Teachers don't have to be masters of the technology - letting students troubleshoot and help each other is a huge burden off the back of a teacher if they are willing to trust their students."*

**OPPORTUNITIES FOR ADDITIONAL RESEARCH**

Given additional time and resources, it would be interesting to conduct some follow-up studies based on this initial research:

- Conducting a student survey against a broader range of students might yield some additional and more diverse insights into the needs and interests of students using PBL in their classes.
- Observations on the use of technology in practice might allow for a richer classification of technology tools, and might provide additional insights into the myriad of ways students and teachers use these technology tools in practice.

**ACKNOWLEDGEMENTS**
The author wishes to thank the students and teachers who took the time to participate in this research study, as well as friends and colleagues who were willing to reach out to their personal and professional contacts to recruit survey participants.

**REFERENCES**

[1] Campbell, C. Problem-based learning and project-based learning. *Australian Council for Educational Research - ACER*, 2014. https://www.teachermagazine.com.au/articles/problem-based-learning-and-project-based-learning.

[2] Corcoran, B. When Tech Meets Project-Based Learning - EdSurge News. *EdSurge*, 2016. https://www.edsurge.com/news/2016-06-22-when-tech-meets-project-based-learning.

[3] Darling-Hammond, L., Zielezinski, M. and Goldman, S. Using Technology to Support At-Risk Students' Learning. *SCOPE: Stanford Center for Opportunity Policy in Education*, 2017. https://edpolicy.stanford.edu/sites/default/files/scope-pub-using-technology-report.pdf.

[4] Dwyer, D. Apple Classrooms of Tomorrow: What We've Learned. *Educational Leadership 51*, 7 (1994).

[5] Hernandez, M. Evaluation Within Project-Based Learning. *Edutopia*, 2016. https://www.edutopia.org/blog/evaluating-pbl-michael-hernandez.

[6] Laan, K. 11 Essential Tools for Better Project-Based Learning. *TeachThought*, 2017. http://www.teachthought.com/project-based-learning/11-tools-for-better-project-based-learning/.

[7] Larmer, J. and Mergendoller, J. Gold Standard PBL: Project Based Teaching Practices. *Bie.org*, 2015. http://www.bie.org/blog/gold_standard_pbl_project_based_teaching_practices.

[8] Lenz, B. and Kingston, S. 17 Teacher Tech Tools for High Quality Project-Based Learning - Getting Smart by Guest Author - Blog, PBL. *Getting Smart*, 2016. http://www.gettingsmart.com/2016/01/17-teacher-tech-tools-for-high-quality-project-based-learning/.

[9] Liebtag, E. 14 Tech Tools to Enhance Project-Based Learning. *Getting Smart*, 2016. http://www.gettingsmart.com/2016/06/14-tech-tools-to-enhance-project-based-learning/.

[10] Rowan, B. 5 Useful Free Web Tools for Project Based Learning Assignments | Emerging Education Technologies. *Emergingedtech.com*, 2013. http://www.emergingedtech.com/2013/11/5-useful-free-web-tools-for-project-based-learning-assignments/.

[11] Sheehy, K. Tips for Transitioning to Project-Based Learning. *U.S. News and World Report*, 2013. https://www.usnews.com/education/blogs/high-school-notes/2013/06/24/tips-for-transitioning-to-project-based-learning.

[12] Thomas, J. *A Review of Research on Project-Based Learning*. Buck Institute for Education, 2000.

[13] Tyson, K. 15 Digital Tools that Support Project-Based Learning l Dr. Kimberly's Literacy Blog. *Dr. Kimberly's Literacy Blog*, 2013. http://www.learningunlimitedllc.com/2013/01/15-tools-that-support-project-based-learning/.

[14] Zielezinski, M. What a Decade of Education Research Tells Us about Technology in the Hands of Underserved Students. *Ed.stanford.edu*, 2016. https://ed.stanford.edu/news/what-decade-education-research-tells-us-about-technology-hands-underserved-students.